\documentclass[10pt]{iopart}
\usepackage[usenames,dvipsnames]{color}
\usepackage[]{graphicx}
\usepackage{placeins}
\usepackage{bm}

\usepackage[]{hyperref}
\hypersetup{
			colorlinks=true,
			linkcolor=blue,
			anchorcolor=black,
			citecolor=blue,
			urlcolor=blue,
}
\usepackage{natbib}
\usepackage{har2nat}   
\setcitestyle{aysep={,}} 

\begin{document}

\title[Turbulence in the solar wind, Voyager 2 data]{Turbulence in the solar wind: spectra from Voyager 2 data at 5 AU}

\author{F Fraternale$^1$, L Gallana$^1$, M Iovieno$^1$, M Opher$^2$, J D Richardson$^3$ and D Tordella$^1$}

\address{$^1$ Dipartimento di Ingegneria Meccanica e Aerospaziale, Politecnico di Torino, Corso Duca degli Abruzzi 24, 10129 Torino, Italy}
\address{$^2$ Astronomy Department, 
   Boston University, 725 Commonwealth Avenue, Boston, MA 02215, USA}
\address{$^3$ Kavli Institute for Astrophysics and Space Research, Massachusetts Institute of Technology (MIT), 77 Massachusetts Avenue, Cambridge, MA 02139, USA}
\ead{daniela.tordella@polito.it}
\vspace{10pt}
\begin{indented}
\item[]
\end{indented}

\begin{abstract}
\textcolor{black}{Fluctuations in the flow velocity and magnetic fields are ubiquitous in the Solar System.  
These fluctuations are turbulent, in the sense that they are disordered and span a broad
range of scales in both space and time. The study of solar wind turbulence is motivated by a
number of factors all keys to the understanding of the Solar Wind origin and thermodynamics. }

The solar wind spectral properties are far from uniformity and evolve with the increasing distance from the sun. Most of the available spectra of solar wind turbulence  were computed at 1 astronomical unit, while accurate spectra on wide frequency ranges at larger distances are still few. In this paper we consider solar wind spectra derived from the data recorded by the Voyager 2 mission during 1979 at about 5 AU from the sun. 
Voyager 2 data are an incomplete time series with a voids/signal ratio that typically increases as the spacecraft moves away from the sun (45\% missing data in 1979), making the analysis challenging. In order to estimate the uncertainty of the spectral slopes, different methods are tested on synthetic turbulence signals with the same gap distribution as V2 data.
Spectra of all variables show a power law scaling with exponents between -2.1 and -1.1, depending on frequency subranges. Probability density functions (PDFs) and correlations indicate that the flow has a significant intermittency. 
\end{abstract}

%
\vspace{2pc}
\vskip10pt
Published in: Physica Scripta 91(2), 023011 (2016).\url{http://dx.doi.org/10.1088/0031-8949/91/2/023011}.

%

%
%
%

\section{Introduction}

\textcolor{black}{Throughout history, it has been human curiosity driven by the need for new knowledges and new resources that pushed humanity to explore. Looking for a very recent example? Consider the recent signed  SPACE Act of 2015 Passed in the US House (H.R. 2262) dealing with the future mission to extract valuable resources from asteroids to fuel an off-Earth economy. Much like the pioneers before us searched for new frontiers to explore and expand into. }

\textcolor{black}{A need that we must learn to meet to safely explore and surf the space is the reconstruction and interpretation of fluctuating signals coming from spacecrafts which for a number of different reasons are usually non regularly acquired on Earth or on other communicating spacecrafts (National Research Council, Decadal Strategy for Solar and Space Physics, 2013)\nocite{nrc-decadal}.}

\textcolor{black}{The Solar System is our nearby garden and is breezed by the Solar Wind which in turn host a swarmth of waves and turbulence launched by the Sun during his varied activities. }
\textcolor{black}{This work focuses on turbulence fluctuations in the solar wind plasma and magnetic field in the ecliptic plane, in the outer solar system at about 5 astronomical units (AU) from the sun. The solar wind fills the heliosphere from the Sun to the termination shock with a supersonic flow of magnetized plasma. This flow is time-dependent on all scales and expands with distance. The solar wind is characterized by a broad range of phenomena, in particular sharp changes in the flow and extreme conditions can often be met due to the crossing of the heliospheric current sheet, the presence of shocks and interaction regions between slow and fast wind streams and zones of strong density variations}. Flow  fluctuations are not just convected outward but show active energy cascades among the different scales. The solar wind turbulence phenomenology has been comprehensively reviewed  by \citet{tu1995} and \citet{bruno2013}.  

Most studies of solar wind turbulence use data from near-Earth, with spacecrafts in the ecliptic near 1 AU \citep[see]{tu1995}.  Recent studies of the solar wind near 1 AU found the fluctuations in magnetic field are fit by power laws with exponents of -5/3 while those of velocity often show exponents of -3/2  \citep{podesta2007}. The Ulysses spacecraft provided the first observations of turbulence near the solar polar regions \citep[see][]{horbury2001};  hourly-average Ulysses data show that the velocity power law exponent evolves toward -5/3 with distance from the Sun, and that spectra at 1 AU are far from the asymptotic state \citep{roberts2010}.   
In order to understand the evolution of the solar wind and its properties, it is necessary to analyze data at large radial distances. However, data gaps typically increase with the distance and make the spectral analysis challenging.
 In this paper we use  Voyager 2 (V2) plasma and magnetic field data from near 5 AU \textcolor{black}{with the aim of obtaining the power density spectra over five frequency decades. Given the Voyager sampling rate (96 seconds for the plasma velocity and 48 seconds for the magnetic field), this might be only possible if one observes a 6-months time interval.} 
 Voyager 2 was launched in August 23, 1977 and reached a distance of 5 AU in the first half of 1979 (just before the Jupiter fly-by - V2 closest approach to Jupiter was on July 9). We use data from January 1 to June 29, 1979, day-of-year (DOY) $1-180$. That year V2 sampled plasma data with a resolution of 12 seconds. After that, telemetry constraints decreased the data sampling first to  every 96 seconds, and now every 192 seconds.
 The Voyager plasma experiment observes plasma currents in the energy/charge range $10-5950\ eV/q$ using four modulated-grid Faraday cup detectors \citep{bridge1977}. The observed currents are fit to convected isotropic proton Maxwellian distributions to derive the parameters (velocity, density, and temperature) used in this work. Magnetic field and plasma data are from the COHO web site (http://omniweb.gsfc.nasa.gov/coho/).
 
In 1979 data gaps are due mainly to tracking gaps; some smaller gaps are due to interference from other instruments. As a consequence, datasets from Voyager 2 are lacunous and irregularly distributed. In order to perform spectral analysis, methods for signal reconstruction of missing data should be implemented. 
 
The paper is organized as follows. \textcolor{black}{In section two, we give an overview of the data recorded around 5 AU by V2, in particular we give information on the average quantities and other statistics of the solar wind physical variables. In section three, three signal reconstruction methods are presented. The methods are then verified by means of a numerically  obtained time series which is a superposition of harmonic oscillations with random phases and a prescribed spectrum. This time series is spoiled by introducing the same sequence of gaps present  in the Voyager 2 data. 
In section four, the plasma and magnetic field spectrum of both the original gapped signals and  reconstructed datasets are shown and discussed through the evaluation of the spectral power scaling in the inertial range. Conclusions and future developments follow in section five.}

\section{Solar wind at 5 AU}
\begin{table*}
\lineup
\caption{\label{parameter} Voyager 2 average quantities for the period 1979, days-of-year [1-180]. }
\begin{indented}
\centering
		\item[]\begin{tabular}{@{}llll}
			\br
		\centre{2}{Parameter} & \centre{2}{Value}                      \\
			\mr
			\boldmath $V_{SW}$        & Mean velocity                        & $4.54\cdot10^2$  & km/s         \\
			\boldmath $V_{A}$           & Alfv\'{e}n velocity                      & $4.94\cdot10^1$            & km/s         \\
			\boldmath $E_v$             & Kinetic energy                       & $1.20\cdot10^3$  & km$^2$/s$^2$ \\
			\boldmath $E_m$             & Magnetic energy                      & $1.37\cdot10^3$  & km$^2$/s$^2$ \\
			\boldmath $E$                & Total energy                         & $2.57\cdot10^3$  & km$^2$/s$^2$ \\
			\boldmath $H_c$             & Cross helicity                       & 15.8              & km$^2$/s$^2$ \\
			\boldmath$L_{E_v}$          & Kinetic correlation length           & $3.68\cdot10^7$  & km           \\
			\boldmath$L_{E_m}$          & Magnetic correlation length          & $3.75\cdot10^7$  & km           \\
			\boldmath${\lambda_v}$    & Kinetic Taylor scale                 & $2.93\cdot10^7$  & km           \\
			\boldmath${\lambda_m}$    & Magnetic Taylor scale                & $2.11\cdot10^7$  & km           \\
			\boldmath $ n_{i}$          & Numerical density                    & 0.23             & cm$^{-3}$    \\
			\boldmath $ E_T$            & Ions thermal energy                       & 2.29             & eV           \\
			\boldmath $ T$              & Ions temperature                          & $2.70\cdot10^4$         & K            \\
			\boldmath $\beta_{p}$       & Ions plasma beta                          & 0.22            &  \\
			\boldmath $c_{s}$           & Ions sound speed                          & $1.93\cdot10^1$          & km/s         \\
			\boldmath $f_{ci}$          & Ions Larmor frequency                & 0.02            & Hz           \\
			\boldmath $f_{pi}$          & Ions plasma frequency                & 0.10              & kHz           \\
			\boldmath $f*$              & Convective Larmor frequency          & 0.44             & Hz           \\
			\boldmath $r_{ci}$          & Ions Larmor radius                        & $4.29\cdot10^3$             & km           \\
			\boldmath $r_{i}$           & Ion inertial radius                 & $1.58\cdot 10^2$              & km           \\
			\br                         &
		\end{tabular}
\end{indented}
\end{table*}

The data set we use is from January 1, 1979 to June 29, 1979 and the RTN Heliographic reference system is adopted. \textcolor{black}{The RTN system is spacecraft-centered, with radial (R) axis directed radially outward from the Sun. The tangential (T) axis is given by the cross product of the Sun rotation axis, northward oriented, and the R axis. The normal (N) axis completes the right handed set}. Figure 1 shows the fluctuations of each component for this period, where the use of Alfv\'{e}nic units  $\bm b=\bm B/\sqrt{4\pi \rho}$  ($\rho$ is the ions density) for the magnetic quantities allows  direct comparison with the velocity field $\bm V$. Higher fluctuations of velocity can be observed in the radial direction, while in the other two directions the magnetic fluctuations are slightly larger. In the same plot it is also shown a zoom focusing on only 4 days, which shows the sections where data are missing.
\textcolor{black}{In the period considered the solar wind has a mean velocity of 454 km/s, but the wind has been subject to shocks and interactions between slow and fast wind streams at smaller radial distance from the sun.} In the low-speed wind the fluctuations are generally higher than in the fast wind, and the magnetic energy is generally higher than  the kinetic energy \citep{marsch1990a,mccomas1998}. 
In fact, the energy per mass unit is about 15\% higher in the magnetic field with respect to the plasma velocity, while the integral scales are almost equal, as shown in table \ref{parameter}. In the same table other characteristic parameters and frequencies are reported. Note that the normalized cross-helicity $\sigma_c=2H_c/E=2\langle\delta \bm V\cdot \bm b\rangle/\langle {\delta \bm V}^2+{\delta \bm b}^2\rangle$ is low ($\approx 0.01$). 
\begin{figure*}
\centering
\includegraphics[width=.8\textwidth]{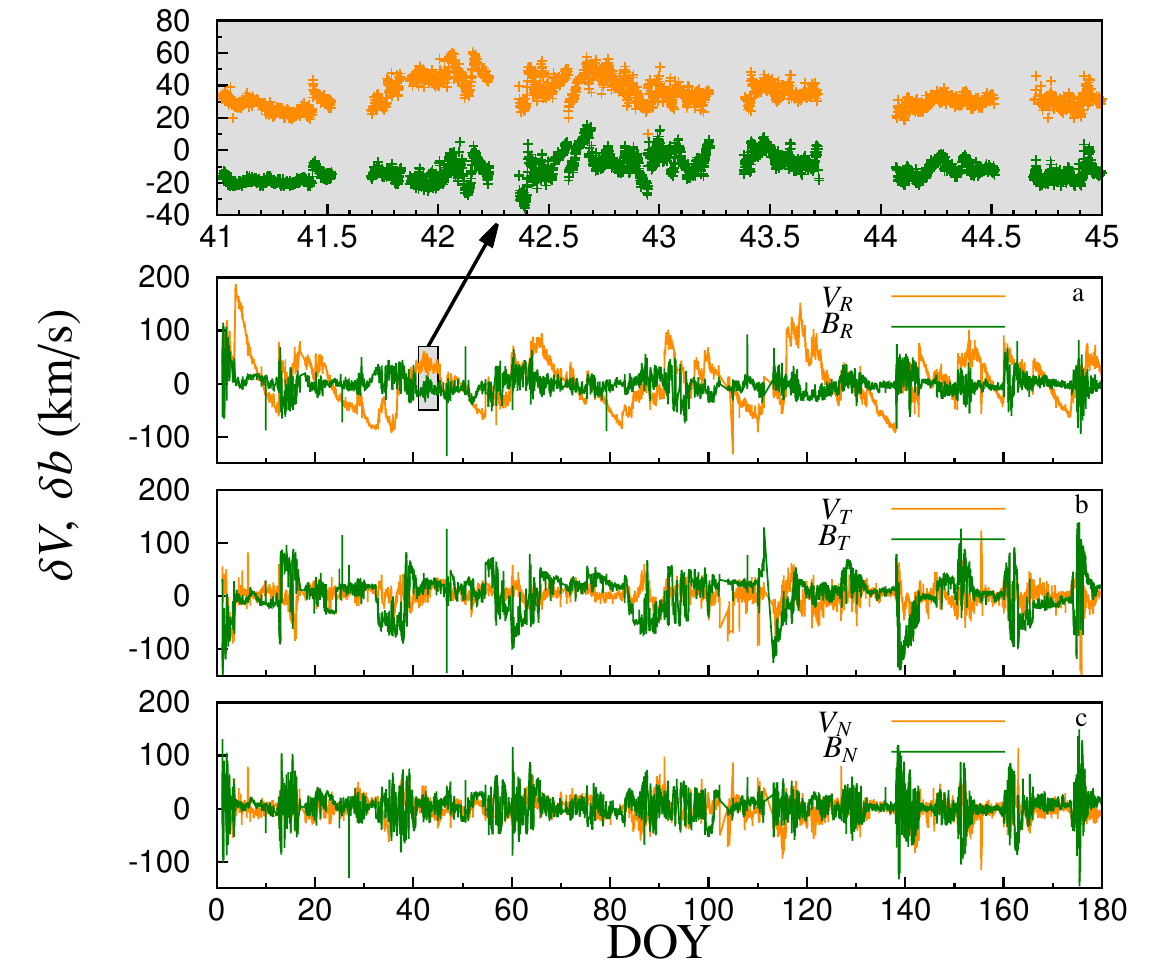}
\label{fig:data}
\caption{Time evolution of velocity (orange lines) and magnetic (green lines) fluctuations in the RTN reference system. The magnetic field is represented in Alfv\'{e}nic units. ({\bf a}) Radial velocity fluctuations, $\delta V=V-\langle V\rangle$, and radial magnetic field fluctuations, $\delta b=b-\langle b\rangle$; ({\bf b}) Tangential fluctuations components; ({\bf c}) Normal fluctuations components.
In the top panel, a zoom of a four day period is highlighted, in which the presence of the gaps can be perceived. The low-speed wind shows a low degree of Alfv\'{e}nic correlation.}
\end{figure*}
\begin{table}
\lineup
\caption{\label{tab:stat}  First four statistical moments for the velocity and magnetic fields normalized modules and for their components. $\mu$ is the mean value, $\sigma^2$ the variance, $Sk$ the skewness and $Ku$ the kurtosis. Notice that the modules of the fluctuations (first two table lines) are computed from the standard-normalized vector components, see  Eq. \ref{eq:normalization} in the text below, so all their statistical moments are dimensionless. In regard to the fields components, velocity units are $km/s$ and $km^2/s^2$ for mean and variance respectively. The magnetic field units are $nT$ and $nT^2$. Skewness and Kurtosis are dimensionless. }
\begin{indented}
\centering
		\item[]\begin{tabular}{@{}lllll}
		\br
		\textbf{} & \boldmath $\mu$&\boldmath $\sigma^{2}$&\boldmath $Sk$ & \boldmath $Ku$\\ 
		\mr
		\boldmath $|V^*|^2$&   3.00      &  10.47      &  2.40   &  10.27\\
		\boldmath $|B^*|^2$&  2.48     &  17.41     &  3.17   &  14.90\\
		\mr  
		\boldmath $V_{R}$&   454      &  1893      &  0.43   &  3.41\\ 
		\boldmath $V_{T}$&   3.21     &  252.9     & -0.99   &  7.35\\ 
		\boldmath $V_{N}$&   0.51     &  250.3     & -0.36   &  5.80\\
		\boldmath $B_{R}$&  -0.04     &  0.173     &  0.53   &  6.71\\ 
		\boldmath $B_{T}$&   0.06     &  0.85      &  -0.72  &  10.2\\ 
		\boldmath $B_{N}$&   0.10     &  0.34      &  -0.24  &  7.65\\
		\br
	\end{tabular}	
\end{indented}
\end{table}

\begin{figure*}
\begin{minipage}{.49\textwidth}
\centering
\includegraphics[width=.9\textwidth]{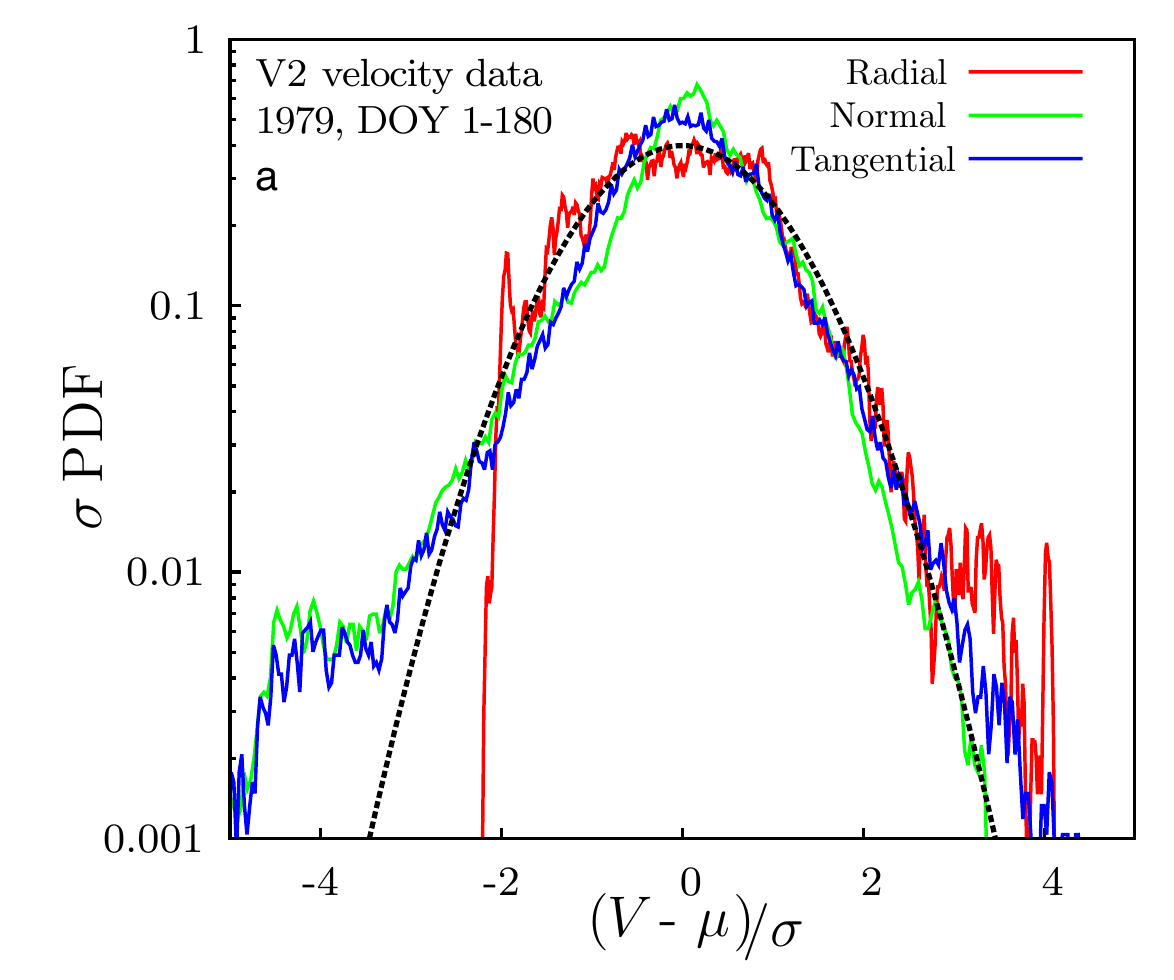}
\end{minipage}
\begin{minipage}{.49\textwidth}
\centering
\includegraphics[width=.9\textwidth]{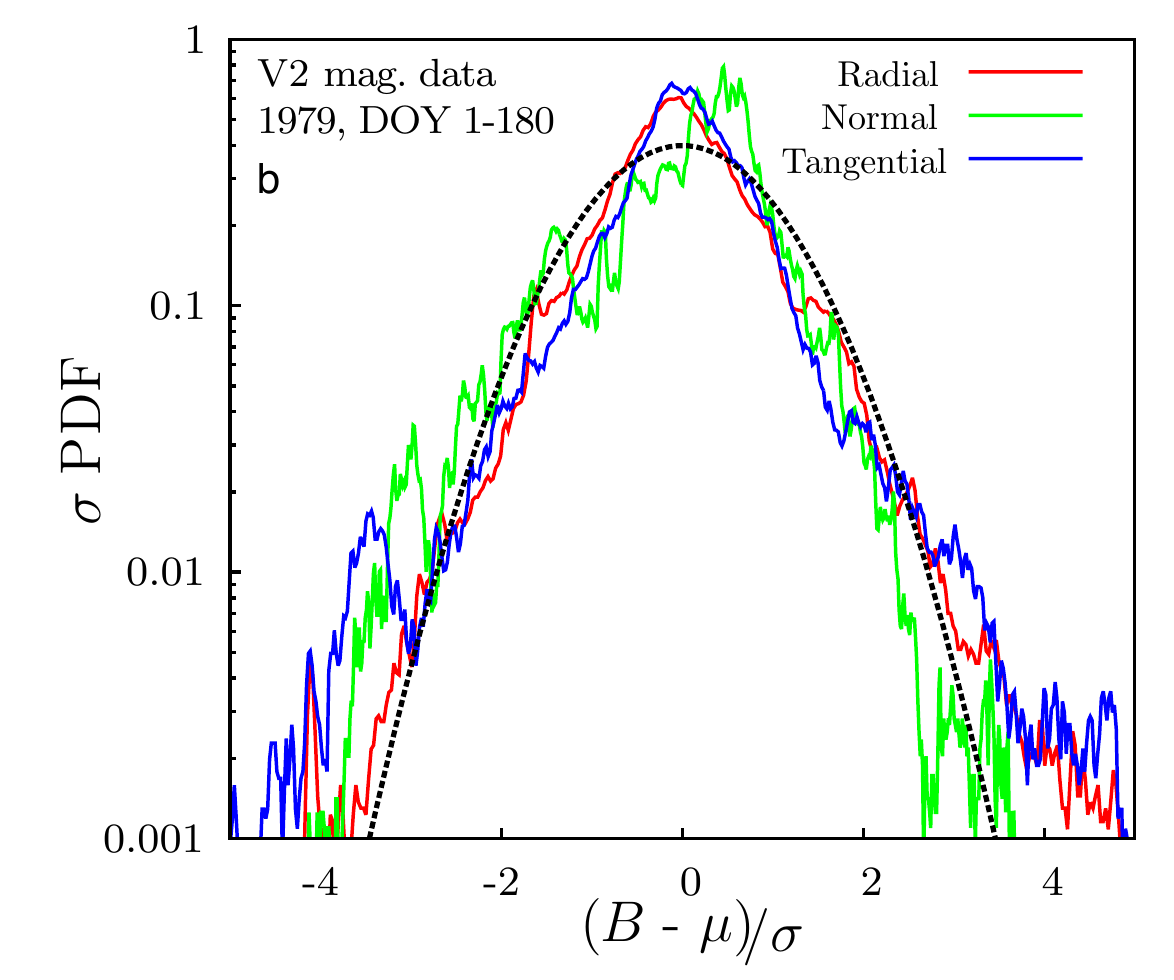}
\end{minipage}
\begin{minipage}{.45\textwidth}
\centering
\includegraphics[width=.9\textwidth]{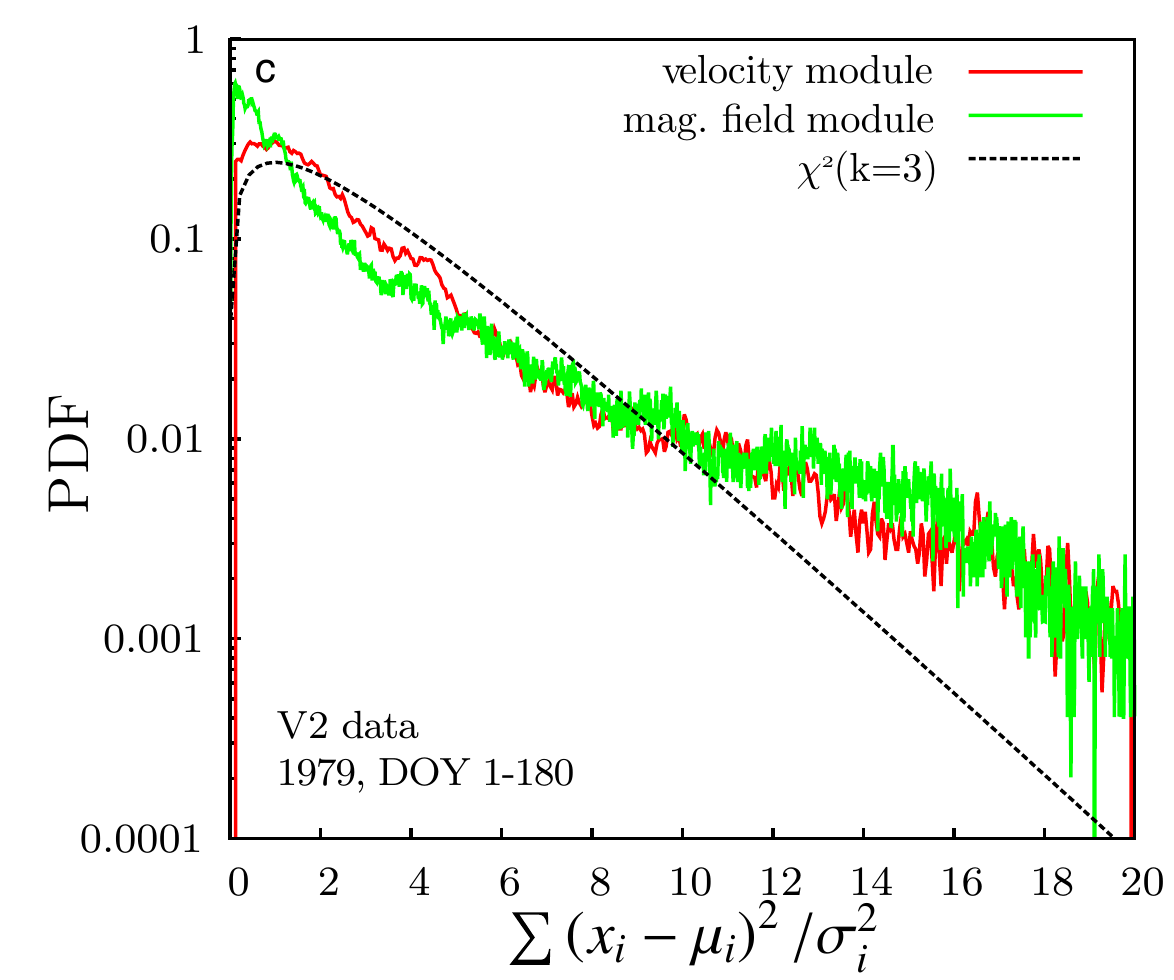}
\end{minipage}
\hfill
\begin{minipage}{.49\textwidth}
\label{fig:pdf}
\caption{Normalized probability density functions (PDF) of the velocity and magnetic field components in the RTN reference system: the velocity is represented in panel (\textbf{a}) and magnetic field in panel (\textbf{b}); both are compared with a normal distribution (black line). Panel (\textbf{c}) presents the normalized PDFs of velocity and magnetic fluctuations modules compared with a 3-component ($k=3$) chi-square distribution (black line).}
\end{minipage}
\end{figure*}

The presence of intermittency in the velocity and magnetic fields can be observed by looking at the probability density functions (PDFs) of the modules of the vector fields, shown in figure \ref{fig:pdf} ({\bf c}). The same plot shows a  three-component chi-square distribution as a reference. \textcolor{black}{Notice that in order to make this comparison, the modules of the velocity, $|\mathbf{V*}|$, and the magnetic field, $|\mathbf{B*}|$ , are computed from the standard-normalized vector components: 
	\begin{equation}\label{eq:normalization}
     	|\mathbf{V*}|^2=\sum_{i=1}^3 \frac{(v_i-\mu_{v_i})^2}{\sigma_{v_i}^2}\hspace{10pt} |\mathbf{B*}|^2=\sum_{i=1}^3 \frac{(B_i-\mu_{B_i})^2}{\sigma_{B_i}^2},
	\end{equation}
 where $\mu$ is the average and $\sigma$ is the standard deviation. The first four moments for these normalized modules are shown in the first two lines of table \ref{tab:stat}}.
Intermittency occurs over a broad range of scales and seems to be slightly higher in the magnetic field data which has larger skewness $Sk$, and kurtosis $Ku$ (see table \ref{tab:stat}). Anisotropy of the fields can be observed by looking at the single components in figure \ref{fig:pdf}({\bf a, b}) . Particularly important are the differences of the radial components compared to the tangential and normal ones: A quantification of the anisotropy can be appreciated by comparing the skewness values in table \ref{tab:stat}.

\section{Spectral analysis of lacunous data: methods and validation with synthetic turbulence data}


\begin{table}
\lineup
\caption{\label{tab:data_gaps} Information on  time intervals among plasma and magnetic field data of Voyager 2, for the period Jan 1-Jun 29, 1979. $\delta t$ represents a time-difference between two consecutive data points, $\delta t_s$ is the basic data resolution. Angle brackets indicate the ensemble average operator. The total number of samples is $n$. $Ls$ indicates the length of a continuous data segment, i.e.  a subset with no missing points with respect to the basic sampling rate. }
                \begin{tabular}{@{}llllllll}
		\br
		\textbf{Jan 1 - Jun 29} &  $n$ & $\delta t_s$ (s)& $\delta t_{min}$ (s) &  $\delta t_{max}$ (hrs) &  $\langle\delta t\rangle$ (s)& $Ls_{max}$ (hrs)&$missing \; data$\\ 
		\mr	
		plasma &$115102$ &    $96$&  $9.6$&$44.7$& $134$ &$19.8$ (at doy $176$) &$28\%$\\
		magnetc field &$248159$&  $48$&  $4.8$&$44.6$& $63$   &$19.5$ (at doy $168$)&$24\%$\\	
		\br
	\end{tabular}	
\end{table}

The data from the Voyagers  suffer from increasing sparsity as the probes move outward in the solar system. There are many causes of data sparsity, noise and artificial unsteadiness in the signals, the most important of which are: (i) tracking gaps due to the V2 location and due to limited deep space network availability; (ii)  interference from other instruments; (iii) possible errors in the measurement chain (from the Faraday cups up to the data acquisition system and the signal shipping to Earth); (iv) the temporal sequence of the propulsion pulses (nuclear propulsion) used to control the Voyager trajectory and thrusters pulses on the  spacecraft used to assist in several critical repositionings of the spacecraft.

\begin{figure*}
\centering
\includegraphics[width=.7\textwidth]{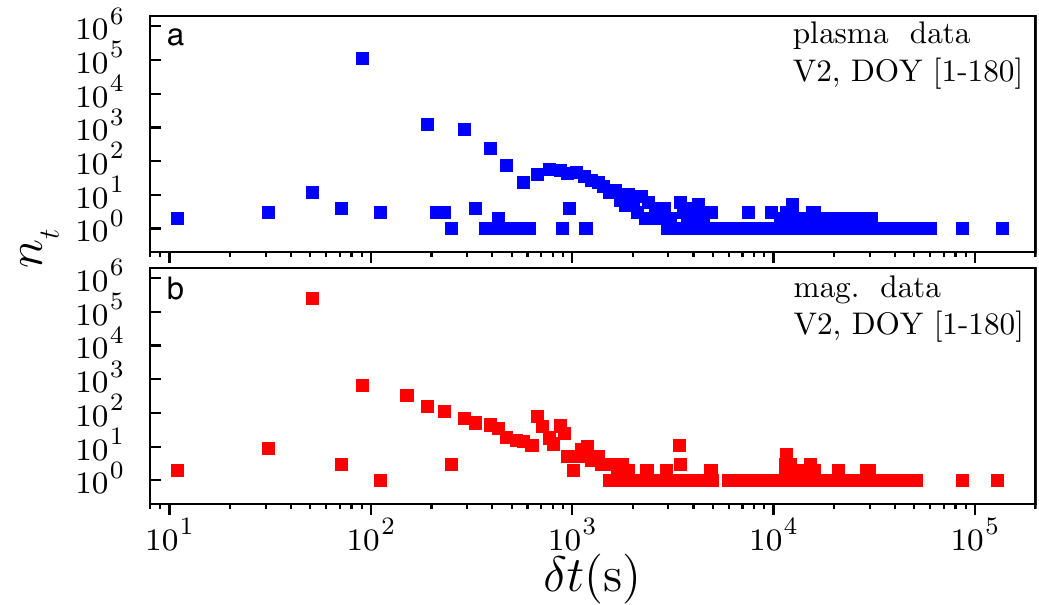}
\caption {Distribution of the time interval between consecutive V2 data points (period 1979, DOY 1-180). \textcolor{black}{$n_t$ is the number of data points spaced by a time interval $\delta t$, and it is traced as a function of the interval length $\delta t$ }. ({\bf a})  Plasma data (i.e. velocity, density, and thermal speed data). The sampling frequency is $\delta t_{sV}=96$ s, so this value is the most frequent among the data.  ({\bf b})  Magnetic field data. The data resolution here is $\delta t_{sB}=48$ s. Note that in both cases many time intervals $\delta t=\delta t_{s}+j\cdot\delta t_{s}$, with $ j=1,2,...$, characterize the data. Note also that $\delta t=\delta t_{s}+j\cdot\delta t_{s}$ represents a gap of ($j-1$) missing data.}
\label{fig:gaps_dist}
\end{figure*}

We will use the following notation:  $\delta t$ is the time interval between two consecutive data points; $\delta t_s$ is the resolution of the data used, or the sampling rate of the instruments; $Ls$ length of a data subset, or segment; $Lg$ is the length of the data gap. 
 In the period considered, the resolution was 96 s and 48 s for plasma and magnetic field data, respectively. For the Voyager magnetometer experiment, the 48 s data comes from averages of higher resolution data (1.92 s and 9.6 s \textcolor{black}{ - for a review about the Voyager MAG experiment, see \cite{behannon1977})}. The longest continuous (i.e. no missing data, $\delta t\leq\delta t_s$) data subset is $Ls\approx19.5$ hrs. Further information on these datasets are shown in table \ref{tab:data_gaps}, while the distribution of $\delta t$ is shown in figure \ref{fig:gaps_dist}.\\
 
\begin{figure*}[]
\centering
	\includegraphics[width=\textwidth]{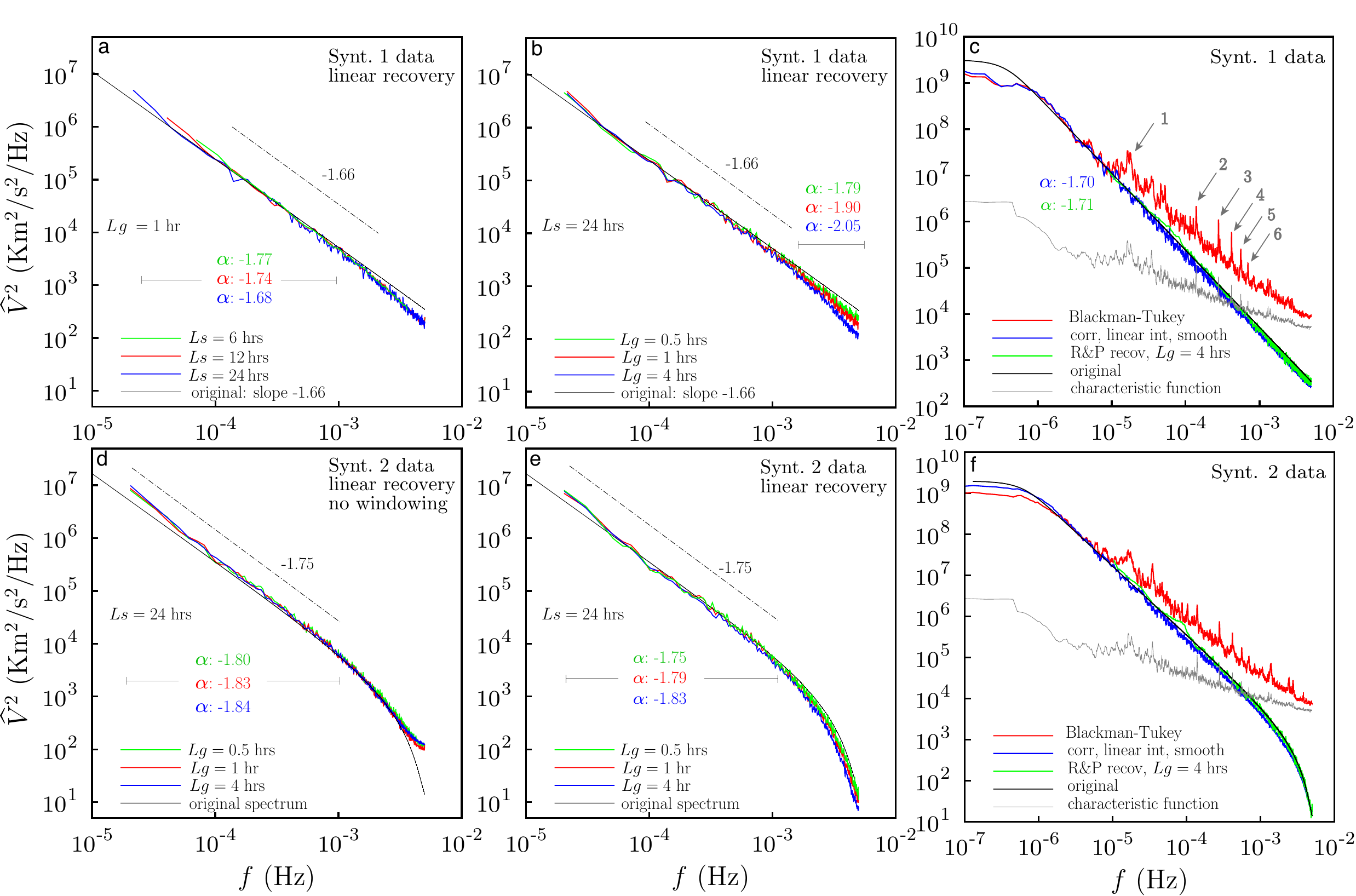}
	\caption {Different methods of spectral analysis are applied to synthetic turbulence data with gaps. $Lg$ is the maximum gap length where interpolation is performed, $Ls$ the length of continuous data segments after interpolation and $\alpha$ is the exponent of the power law fit. The black curves represent the correct spectrum to recover.  ({\bf a},{\bf b}) Averaged spectra from direct FFT (with Hann windowing) after linear interpolation on \textit{Synt1} sequence. For all the averaged spectra shown, the number of subsets used is between 80 and 400.  The low-pass effect increases as $Lg$, and becomes evident for $f>10^{-3}$ Hz where the lower error ($L_{s}=24$ hrs, $L_{g}\leq0.5$ hr) on the slope is around $8\%$. For lower frequencies this error is about $2.8\%$. ({\bf d},{\bf e}) The same method is applied to \textit{Synt2}, and the effect of a windowing technique is shown. ({\bf c},{\bf f}) Red line: spectrum from two-point correlation computed from raw data (Blackman-Tukey method with 10\% taper cosine window). The influence of the gap distribution is evident, if the reader compares the red curve with the grey one, representing the spectrum of gaps. Indeed both  non-physical peaks and $1/f$ noise show up. The frequency distance between peaks 2-6 is constant, $\Delta f_p=0.0014$. Non-physical peaks can be recognized by tracing the spectrum of the function $\Phi$  (grey line). Blue line: correlations computed after linear interpolation of data; the original spectrum is recovered with an error of $2.5\%$ for all frequencies. Green line: maximum likelihood reconstruction \citep[see][]{rybicki1992a}. In this case a discrepancy of about $2.5\%$ is obtained for an extended frequency range with respect to the direct FFT method (indeed, up to 4 hrs gaps are recovered). }
	\label{fig:synt_results}
\end{figure*}\vspace{-10pt}

In this section we show the analysis procedure followed to perform a reliable spectral analysis and to quantify the uncertainty of the results of section 4. Two sequences  of synthetic turbulence (named \textit{Synt1} and \textit{Synt2}) have been generated from imposed three-dimensional power spectrum and random phases:
\begin{itemize}
\item  \textit{Synt 1} $\to$ $E_{3D}(n/n_{0})=\frac{(n/n_{0})^{\beta}}{1+(n/n_{0})^{\alpha+\beta}}$\
\item  \textit{Synt 2}$\to$ $E_{3D}(n/n_{0})=\frac{(n/n_{0})^{\beta}}{1+(n/n_{0})^{\alpha+\beta}}\bigl[1-exp( \frac{n-n_{tot}}{\gamma}+\epsilon  )   \bigr]$
\end{itemize}\vspace{10 pt}
 $\beta=2$, $\alpha=5/3$,  $n_{0}=11$, $\gamma=10^{4}$, $\epsilon=10^{-1}$

 \textit{Synt1} reproduces the energy-injection range and the Kolmogorov inertial range of canonical 3D fluid turbulence, while \textit{Synt2} reproduces also the dissipative part of the spectrum. Both the sequences mimic the variance and the integral scale of V2 data, and 1D spectra (see black curves of figure \ref{fig:synt_results}) are computed from the relation for homogeneous and isotropic turbulence $E_{3D}(f)=-f\ dE_{1D}/df$ \citep[see][Chap. 3]{monin_book} , where $f$ is the frequency. The same distribution of gaps (see figure \ref{fig:gaps_dist}) of the V2 plasma data has been then projected to the synthetic set. 

Different methods to compute the power spectral density have been tested, and the results are shown in figure \ref{fig:synt_results}.  We consider as parameters for this analysis $Lg$, which here represent the maximum  size of gaps ``filled'' by the interpolation, and $Ls$, the length of continuous segments. Results of direct FFT after linear interpolation are shown in figure \ref{fig:synt_results}\textbf{a},\textbf{b},\textbf{d},\textbf{e}, showing averaged spectra (ensembles of 80 to 400 subsets). The low-pass character of the linear interpolator is evident, resulting in an overestimation of the slopes for all cases, which increases with $Lg$.  For $Lg=30$ min  and $Ls=24$ hrs the error of the exponent $\alpha$ is below $3\%$ for  both \textit{Synt1} and \textit{Synt2} in the range  $10^{-5}\leq f\leq 10^{-3}$ Hz. The discrepancy for $f>10^{-3}$ Hz is about $8\%$. A windowing technique (Hann window) is necessary to eliminate noise effect ($\approx 1/f$) due to segmentation.  To preserve the total energy a factor of 2.66 is applied to the squared Fourier coefficients. The effect of the window can be appreciated by comparing panel (\textbf{d}) and (\textbf{e}) of figure  \ref{fig:synt_results}. To extend the analysis to a broader range of frequencies, we computed the correlation spectra (panels \textbf{c},\textbf{f}). The red curve of figures \ref{fig:synt_results} (\textbf{c},\textbf{f}) results from correlations computed directly from available data. \textcolor{black}{This procedure is due to \citet{blackman1958} and it has been widely used for solar wind spectral analysis at about 1 AU, since it allows to greatly overcome the missing data problem when the missing data are less then 10\%.  However, for higher degree of data sparsity the computation of correlations is delicate, since the convergence to the correct values is slow and the gap distribution deeply affects the spectral results. This result (see the red spectrum in panels \textbf{e} and \textbf{f})} progressively underestimates the spectral slope for  $f>10^{-5}$ Hz. The peaks 1-6 are due to the gap distribution of figure \ref{fig:gaps_dist}. Indeed, these peaks are evident in the spectrum of the characteristic function $\Phi$ (see the grey curve in panels \textbf{c,f}), where $\Phi=1$ when the signal is available, $\Phi=0$ elsewhere.
If a linear interpolation of the data is performed, correlations show a much better convergence to the correct values and the error of the slope $\alpha$ is about $2.5\%$ in the inertial range (blue curves). Note that in this case no averaging is possible, but a smoothing technique is applied by averaging frequencies in a bin of varying width around each point (energy is preserved). 

The last method we test is the maximum likelihood reconstruction described by \citet{rybicki1992a} and applied by \citet{rybicki1992b}. The method is based on a minimum-variance recovery with a stochastic component.  It allows us to compute different statistically equivalent reconstructions, which are constrained by the data when these are available. When the data are missing, the reconstruction consists of the minimum variance interpolation plus a stochastic process based on a correlation matrix.  We compute this matrix from the actual data two-point correlations. Afterwards, averaged spectra of recovered subsets are computed. We see that in this case the accuracy of $2.5\%$ can be achieved by filling up to 4 hrs of gaps (see green curves of figure \ref{fig:synt_results} \textbf{c},\textbf{f}). \par
\textcolor{black}{Typically, the fraction of missing data increase as the distance from the Earth, and differences among the different procedures become more pronounced. Computing averaged spectra from small subsets allows to estimate well the high-frequency range of the spectrum but only two decades can be observed. The correlation method suffers from leakage at the high frequencies due to the low-pass effect of the linear interpolator, this could lead to an overestimation of spectral slopes in case of data with higher amount of missing points. This is compensated by the maximum likelihood recovery method, which interpolates in the gaps according to an input correlation function. The correlation function can be either a model or an estimate from the actual data. In this regard, the first method (averaged spectra of small continuous subsets) may be quite helpful. For instance, for the present data it yields reliable power spectra in the range $f=2\cdot10^{-5}\div 5\cdot10^{-3}$ Hz; this means that good two-point correlation functions of about 14 hours of maximum time lag can be obtained from inverse transform. These are sufficient to fill almost all gaps using the maximum likelihood recovery method.}

\begin{figure}
\centering
\includegraphics[width=.6\textwidth]{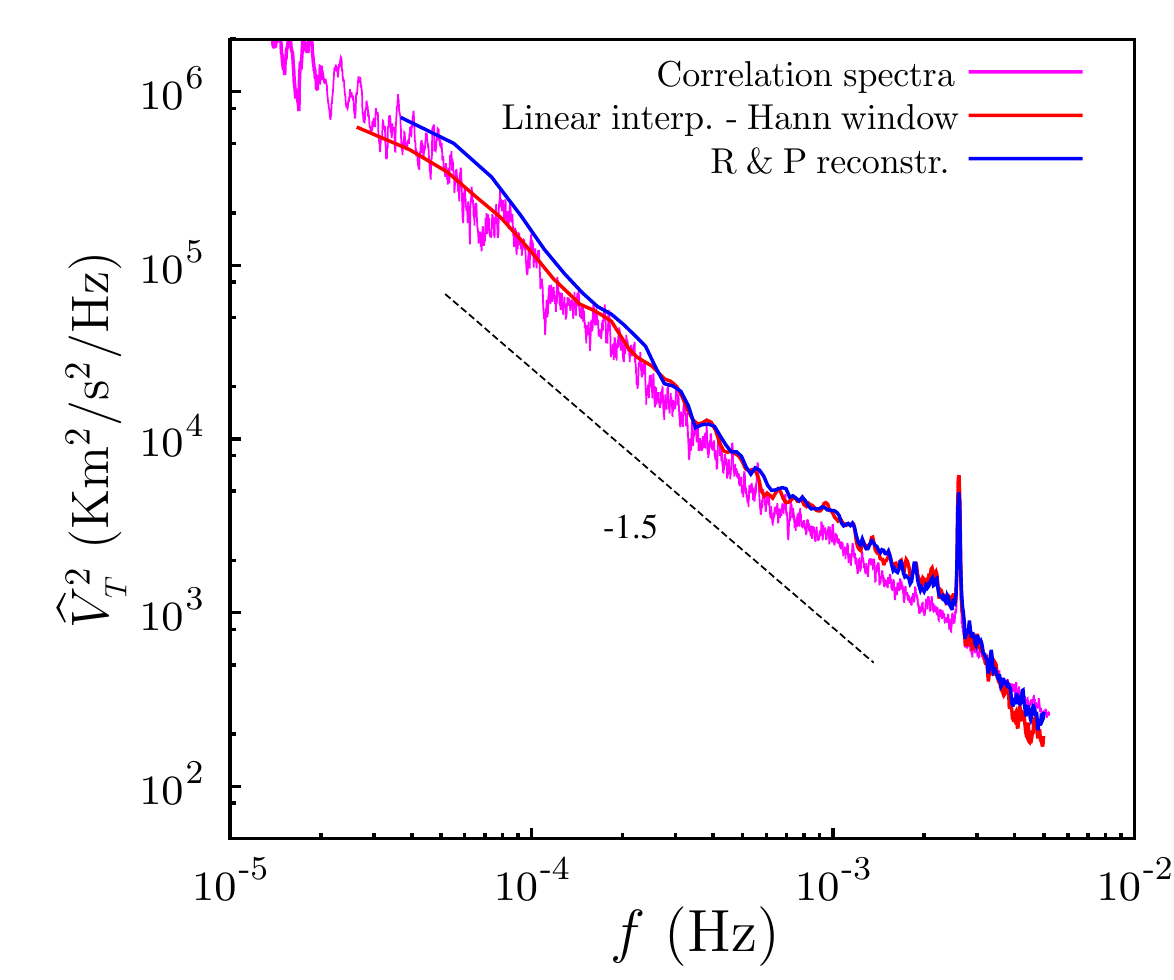}
\caption{\label{fig:spec_comp} \textcolor{black}{Comparison of $V_T$ spectra in the high frequency range (24 hours data subsets) using the three different methods described in Section 3. Spectra by the correlation method with preliminary linear interpolation (pink line) are computed on the 6 months period, the slope in the frequency range [$2\cdot10^-5-10^-3$] Hz is -1.54. The linear interpolation with Hann windowing (red line) consider 24 hours periods. In this case, the largest gap recovered last 30 minutes, and the spectral slope is -1.47.  About the maximum likelihood reconstruction (R \& P Reconstruction, blue line), the largest gaps recovered last 4 hours and the spectral slope is -1.56.} }
\end{figure}

\section{\textcolor{black}{Power law spectra at 5 AU. Phenomenological interpretations and relevant spectral accuracy requirements.} }

\textcolor{black}{Spacecrafts yield power spectra of the velocity and magnetic field in terms of the fluctuation frequency.  The identification of a range in the frequency spectrum for which a power law applies 
allows to characterize magnetohydrodynamic (MHD) turbulence inertial energy transfers. 
It should be noted that the relative difference in the power spectral exponents associated to the various forms that the MHD energy cascade models can take is as low as 10\%. In order to distinguish possible different inertial regimes in terms of the power scaling law exponent, it is useful to have different independent methods able to discriminate spectral decay variations well below 10\%. This is the fundamental motivation which has prompted this work toghether with the concomitant spacecraft data sparsity problem. By applying the Taylor frozen-flow hypothesis,  the frequency can be converted to radial wavenumbers. It should be recalled that the wavenumber space is  
the prefereed space where the turbulence cascade among different physical scales is pictured.  
In this regard, let us here just mention two classical phenomenological interpretation which, although based on different physical hypothesis, yield close values of spectral power law exponent (namely -1.66 versus -1.5, 10\% of relative difference). 
In a supposed existing inertial regime, the energy $E$ and the wavenumber $k$ satisfy the law $E(k)\propto k^\alpha$. The values of $\alpha$ depend  on the way the energy is transferred among the inertial scales. The turbulent energy transfer rate is defined, see e.g. Zhou, Matthaeus and Dmitruk 2004 \cite{zhou2004}, as 
$
        \varepsilon = u_k^2 / \tau_{sp}$,
    where $u_k(k)=(k E(k))^{1/2}$ is the velocity associated to a given eddy, $\tau_{sp}(k)=\tau^2_{nl}(k)/\tau_t(k)$ is the spectral transfer time, $\tau_{nl}(k)=(k u_k)^{-1}$ is the characteristic eddy turnover time, and $\tau_{t}$ the time scale of the triple correlation function. Time $\tau_t$ may depend on any relevant turbulent parameter and the wavenumber $k$, depending on the dominant phenomenology.   
		The strain-dominated MHD turbulence is characterized by the direct interaction among vortexes having similar wavenumbers. Adopting the Kolmogorov's concept of independence of widely separated wavenumbers in the inertial range, the MHD motion is comparable to a classical hydrodynamics motion in which the energy power law has been determined by dimensional analysis \cite{kolmogorov1941a}. In this case the time scale of the transfer function is equivalent to the eddy turnover time, and so the relation between $E$ and $k$ is given by $
        E(k) \sim \varepsilon^{2/3}k^{-5/3}$.  
		When the magnetic energy in sub-inertial wavenumbers exceeds the total energy in the inertial range, the asymptotic inertial-range energy spectrum is instead proportional to $k^{-3/2}$ and displays exact equipartition between magnetic and kinetic energy.This happens in the so called {\it sweep-dominated} motion where large-scale and small-scale eddies interact: in particular, a big vortex sweeps oppositely the small ones, depending on their polarity. In that case, correlations decay with an effective lifetime determined by propagation of structures at the Alfv\'en time scale, $\tau_A (k) = (V_Ak)^{-1}$, see \citet{irosh1963} and \citet{kraich1965}, which yields the -3/2 slope.
In the following,  the methods described in section 3 have been applied to the Voyager 2  datasets in order to estimate spectral power law  exponents in the first semester of 1979.	 Figure \ref{fig:spec_comp} shows the output of the three reconstruction methods on the spectrum of the tangential velocity component of the solar wind in the frequency range corresponding to an observation interval of 24 hours. A set of 80 observations distributed in between days 1 and 180  is considered.  Over the three methods, the arithmetic mean of the spectral slope is -1.52 with an offset of 0.05, which corresponds to an accuracy of 3.3\%.  The offset is mainly due to the contribution of the linear interpolation which is the least  accurate among the three methods, as explaned in Section 3, see figure 4. 			
All the results shown in the following are instead obtained by using the correlation method, which has the same accuracy of the Rybicky-Press method (less than 3\%) but is computationally faster.}

\begin{figure}
 \centering
 \includegraphics[width=.6\textwidth]{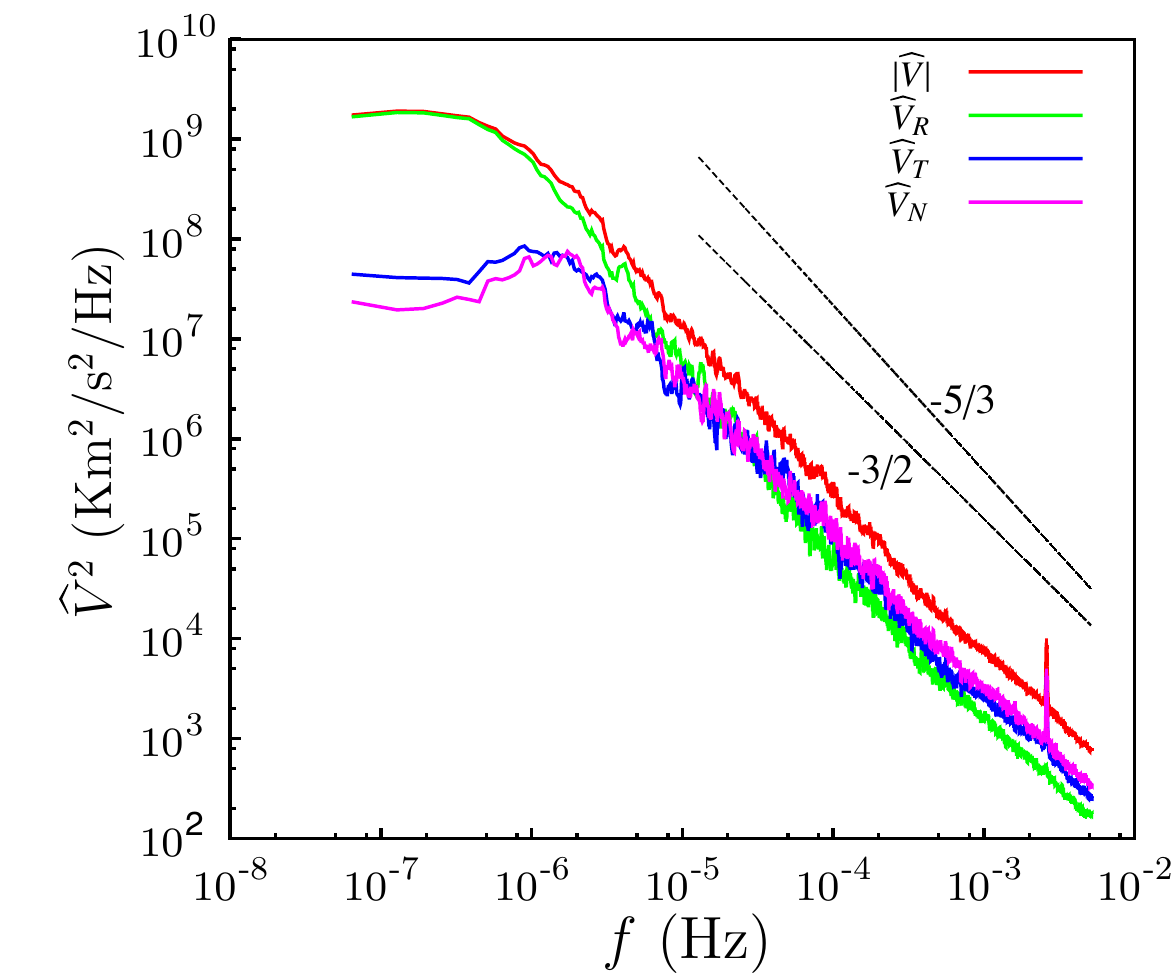}
 \caption{\label{fig:vspectra} \textcolor{black}{ Velocity components and module spectra in the frequency range $6.4\cdot 10^{-8}$ Hz - $10^{-2}$ Hz. Computation by means of the correlation method coupled to a linear interpolation. About the decay power law exponents associated to frequency subranges, see table \ref{tab:data_exponent}.}}
 \end{figure}

In figure \ref{fig:vspectra} the velocity spectra representing frequency content in the six months observation interval  are shown for all the velocity components. To highlight slope variations in the overall six months spectral domain, table \ref{tab:data_exponent} gives for the velocity module, the detail of  the spectral slope estimates on each of the four decades $10^{-6} \div 10^{-2}$ Hz.

\begin{table*}  
\lineup
\caption{\label{tab:data_exponent} Synthesis of the exponents found for velocity and magnetic fields in different frequency ranges, computed using the correlation method coupled with linear interpolation. $b$ is the magnetic field in Afv\'{e}nic units ($\mathbf{b}=\mathbf{B}/\sqrt{(4\pi\rho)}$).
For the velocity field, exponents in the higher frequency range are computed neglecting the peak present at $f=2.6\ mHz$.}
\begin{indented}
\centering
		\item[]\begin{tabular}{@{}llll}
		\br
		$f$ range &  $|\mathbf{V}|$ & $|\mathbf{B}|$ & $|\mathbf{b}|$\\ 
		\mr	
		$10^{-6}-10^{-5}$ & -1.85 & -1.23 & -1.38 \\	
		$10^{-5}-10^{-4}$ & -1.72 & -1.75 & -1.63\\	
		$10^{-4}-10^{-3}$ & -1.26 & -1.75& -1.63\\
		$10^{-3}-10^{-2}$ & -1.35 & -1.93& -1.80\\
		\br
	\end{tabular}	
\end{indented}
\end{table*}


The mean exponent for the kinetic field at frequencies ranging from $10^{-5}$ Hz to $10^{-2}$ Hz, shown in figure \ref{fig:vspectra}, is around $-1.5$. This value is lower than the -5/3 exponent and seems to support the MHD cascade model with the -3/2 Iroshnikov-Kraichnan exponent. The exponent is not constant in the whole range: it starts with higher values ($-1.85$) at low frequencies  and decreases after $f=2\cdot10^{-3}$ Hz, reaching values lower than $-1.35$. As shown by \cite{matt89}, lower exponent are typical of weak turbulence, in which the field is characterized by strong mean values respect to the fluctuations, as in the Iroshnikov and Kraichnan model, where an exponent of $-3/2$ is assumed.  Higher exponents are found for strong turbulence, as in the $-5/3$ Kolmogorov model. The flattening in the velocity spectra at high frequencies has been observed by other authors \citep{matthaeus1982b,roberts2010}. The first authors pointed out that the flattening may be due to aliasing, but the last author excluded this hypothesis. We performed our analysis on synthetic data with the same significant digits  of V2 data to exclude $1/f$ noise. This flattening was also found for the proton density and temperature fluctuations \citep{marsch1990b}, as well as for the Esl\"{a}sser variables \citep{marsch1990a}. The flattening is most typical of high-speed streams below 1 AU, in this cases the frequency range observed is more extended and the change of slope occurs at  $f\approx10^{-7}$ Hz.
The peak in kinetic spectra visible at the frequency $f=2.6\ mHz$ is associated to the data acquisition frequency -- indeed, it is a harmonic of the sampling frequency $f_{s}=10.4\ mHz$. It is not an artifact of the signal reconstruction methods, since  such a peak does not show up in the test cases of figure \ref{fig:synt_results}. Moreover it is most evident in the tangential and normal velocity components, while it is not present in the magnetic field. Analysis of the Voyager 1 data in the same period shows the same harmonic which is not present, for instance, in the year 1978.

\begin{figure}
\centering
\includegraphics[width=.6\textwidth]{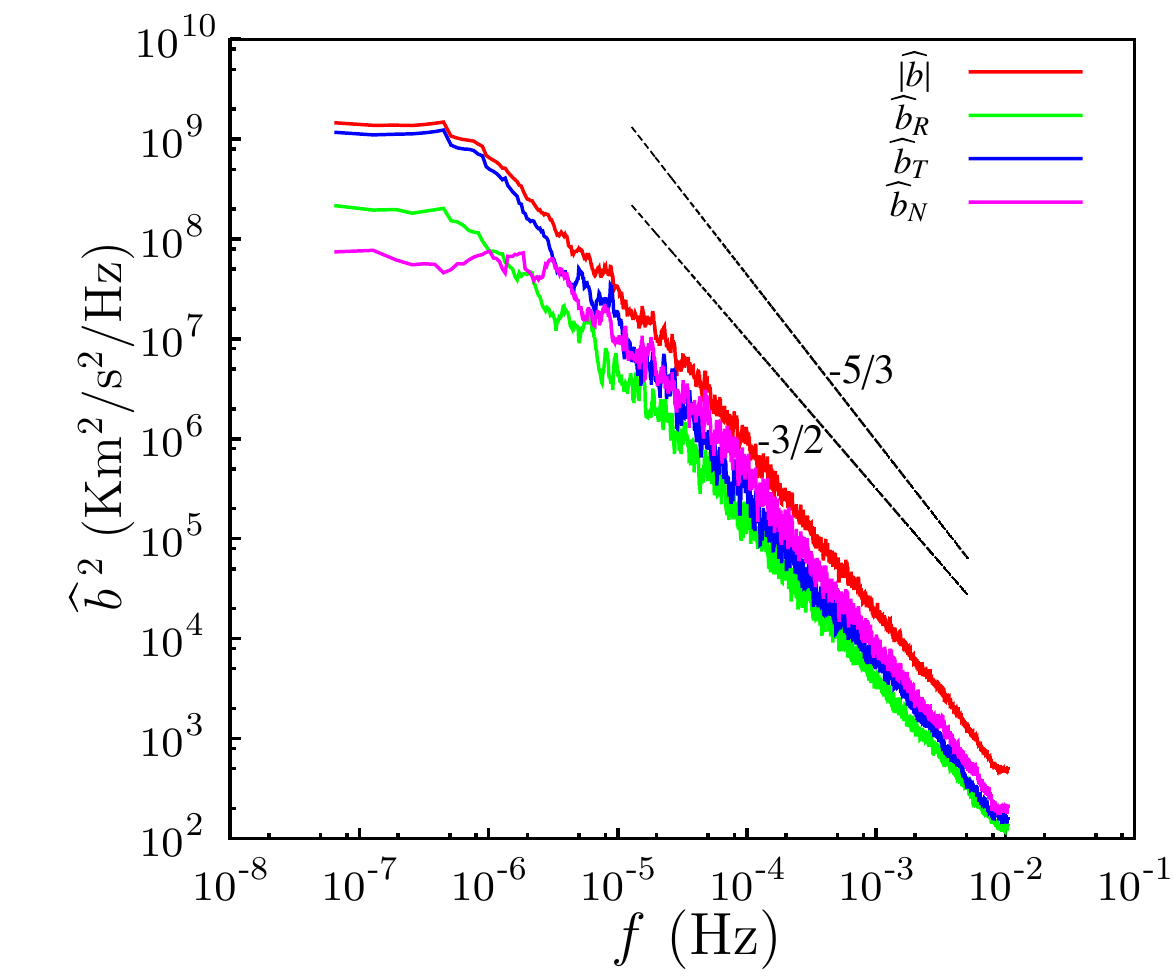}
\caption{\label{fig:bspectra} Spectra of magnetic field components  and module (in Alfv{\'e}n units), in the frequency range $6.4\cdot 10^{-8}$ Hz - $10^{-2}$ Hz. Computation by means of the correlation method coupled to a linear interpolation. About the decay power law exponents associated to frequency subranges, see table \ref{tab:data_exponent}.}
\end{figure}

In the magnetic field spectra shown in figure \ref{fig:bspectra}, the exponents found are higher than those for the velocity and present a $-1.75$ slope in the frequency range $10^{-5}<f<10^{-2}$ Hz.   
At low frequencies the behavior of velocity and magnetic fields is very different: while the exponents for the magnetic spectra decrease to around $-1.2$, the velocity spectra become steeper, reaching values as high as $-1.85$. This high slope for $f<10^-5$ Hz may be due to the presence of jumps or local trends in the signal (mainly in the radial velocity), as shown by \citet{roberts1987c} and \citet{burlaga1989} .
At high frequencies a slight increase in the magnetic slopes is found, with a variation of the order of 10\%, and resulting exponents around $-2$. The $\approx -1$ slope of the magnetic spectra for the low-frequency range has been discussed by \citet{roberts2010}, and it may be related to the presence of large-scale Alfv\'{e}n waves generated at the solar corona. This range typically reduces with the radial distance and for low correlated ($\sigma_{c}\to0$) streams. Though the magnetic field fluctuations in the inertial range often follow a Kolmogorov -5/3 behavior \citep{podesta2007}, spectral indices around 1.8 have been recently observed at about 1 AU by other authors as \citet{safrankova2013}. 

Spectra presented here are in good agreement with those found by  \citet{matthaeus1982b} (magnetic spectral index of -1.7 at high frequencies for Voyager 1 data at about 5 AU) and \citet{klein1992} (magnetic spectral index of -1.17 at low frequencies and -1.88 at high frequencies for Voyager 1 data at 4 AU), where an analysis for V1 spectra of magnetic modules respectively at 5 and 10 AU can be found for frequencies ranges from $10^{-7}$ Hz to $10^{-4}$ Hz.

\section{Conclusions}
In this work we computed the spectra for a frequency range extending over five decades ($6.4\cdot10^{-8}-10^{-2}$ Hz), for the solar wind at 5 AU by recovering the missing data of Voyager 2. 
Three analysis procedures have been validated by testing various reconstruction methods on a well-known turbulent system, hydrodynamic turbulence, in which the same gap distribution as in the V2 plasma data were reproduced. \textcolor{black}{ The estimated uncertainty on the spectral slope is below 3\%,  therefore this allows to reliably distinguish the different scaling laws characterizing the physical fields. These methodologies can in general be applied to other datasets, and the quality of the spectral estimation will depend on the amount and distribution of the missing data along the global observation time interval.}
The exponents of the velocity field are lower than those of the magnetic fields for high frequencies. At low frequencies, however, the velocity spectra became steeper, while the magnetic spectral indices decrease at values lower than $1.5$. 
The results of the analysis are consistent with others in literature, confirming  the validity of the reconstruction techniques used. This outcome allows us to plan the more challenging reconstruction and the consequential spectral analysis of Voyager 2 data (since 2007), where only 3\% of possible data points are available. This analysis would be helpful to better understand the heliosheath plasma and magnetic field behavior. \citep{richardson2008,opher2011}. 
\bibliographystyle{jphysicsB}  
\bibliography{JGR_bibliography_new}

\end{document}